\documentclass{aa}
\usepackage{txfonts,natbib,graphicx}
\bibpunct{(}{)}{;}{a}{}{,}

\newcommand{\irnulzes}{\object{IRAS\,06530-0213}}
\newcommand{\irnulacht}{\object{IRAS\,08143-4406}}
\newcommand{\irnulvier}{\object{IRAS\,04296+3429}}
\newcommand{\irnulvijf}{\object{IRAS\,05341+0852}}
\newcommand{\irnulzeven}{\object{IRAS\,07134+1005}}
\newcommand{\ireennegen}{\object{IRAS\,19500-1709}}
\newcommand{\irtweetwee}{\object{IRAS\,22223+4327}}
\newcommand{\irtweedrie}{\object{IRAS\,23304+6147}}

\newcommand{\pasa}{PASA}
\newcommand{\phscr}{Phys. Scr.}

\defcitealias{vanwinckel00}{Paper~I}

\begin{document}
\title{A study of the s-process in the carbon-rich\\post-AGB stars
IRAS\,06530-0213 and IRAS\,08143-4406\\on the basis of VLT-UVES
spectra\thanks{Based on observations collected at the European Southern
Observatory, Paranal, Chile (ESO Programme 66.D-0171)}$^{\rm ,}$\thanks{Table
\ref{tab:equivalentwdths} is only available in electronic form at the CDS via
anonymous ftp to {\tt\footnotesize cdsarc.u-strasbg.fr (130.79.128.5)} or via
{\tt\footnotesize http://cdsweb.u-strasbg.fr/cgi-bin/qcat?J/A+A/}}}
\author{Maarten Reyniers\inst{1} \and
Hans Van Winckel\inst{1}\thanks{Postdoctoral fellow of the Fund for Scientific Research, Flanders} \and
Roberto Gallino\inst{2,3} \and
Oscar Straniero\inst{4}}
\institute{Instituut voor Sterrenkunde, Departement Natuurkunde en Sterrenkunde,
K.U.Leuven, Celestijnenlaan 200B, 3001 Leuven, Belgium
\and
Dipartimento di Fisica Generale, Universit\'a di Torino, Via Pietro Giuria 1,
10125 Torino, Italy
\and
Centre for Stellar and Planetary Astrophysics, School of Mathematical Sciences,
Monash University, 3800 Australia
\and
Osservatorio Astronomico di Collurania, 64100 Teramo, Italy}
\offprints{Maarten Reyniers, \email{Maarten.Reyniers@ster.kuleuven.ac.be}}
\date{Received 25 September 2003 / Accepted 18 December 2003}

\abstract{In an effort to extend the still limited sample of s-process
enriched post-AGB stars, high-resolution, high signal-to-noise VLT+UVES spectra
of the optical counterparts of the infrared sources \irnulzes\ and \irnulacht\
were analysed. The objects are moderately metal deficient by [Fe/H]\,=\,$-$0.5
and $-$0.4 respectively, carbon-rich and, above all, heavily s-process
enhanced with a [ls/Fe] of 1.8 and 1.5 respectively. Especially the spectrum
of \irnulzes\ is dominated by transitions of s-process species, and therefore
resembling the spectrum of \irnulvijf, the most s-process enriched object
known so far. The two objects are chemically very similar to the 21$\mu$m
objects discussed in \citet{vanwinckel00}. A homogeneous comparison with the
results of these objects reveals that the relation between the third dredge-up
efficiency and the neutron nucleosynthesis efficiency found for the 21$\mu$m
objects, is further strengthened. On the other hand, a detailed comparison
with the predictions of the latest AGB models indicates that the observed
spread in nucleosynthesis efficiency is certainly intrinsic, and proves that
different $^{13}$C pockets are needed for stars with comparable mass and
metallicity to explain their abundances.
}

\authorrunning{M. Reyniers et al.}
\titlerunning{The s-process in IRAS\,06530-0213 and IRAS\,08143-4406}
\maketitle

\section{Introduction}
In the last two decennia, the progress in the theoretical modelling of AGB
stars, as well as the qualitative and quantitative improvement of the
observational data of (post-)AGB stars, are impressive. Whereas, before the
launch of the IRAS satellite, only a few post-AGB candidates were known, the
present sample consists of about 220 objects \citep{szczerba01}. The main
keyword, however, that can be applied to the post-AGB sample is
{\em diversity}. In practically all their aspects, post-AGB stars are much
more diverse than theoretically anticipated.

In the case of the {\em morphology}, the resolved post-AGB stars display a
surprisingly wide variety of shapes and structures. There are only few
sources displaying a spherically symmetric morphology, most sources possessing
a clear bipolar structure \citep[see e.g. ][]{sahai01}. The aspherical
structure was already demonstrated earlier by ground based observations both
in optical images of the scattered visible light
\citep[e.g. ][ R$\sim$0.$^{\prime\prime}$75]{hrivnak99b}, and in mid-IR images
of the thermal emission of dust
\citep[e.g. ][ R$\sim$$1^{\prime\prime}$]{meixner99}. Many more structural
details were, however, resolved on HST images \citep[e.g.][]{ueta00, hrivnak01}.
For a recent review, we refer to \citet{balick02}.

Concerning the {\em circumstellar chemistry}, the high resolution spectroscopy
of the infrared excess also reveals an interesting chemical evolution. The
observations made by the Infrared Space Observatory (ISO) satellite brought
important progress in this domain. In an O-rich chemistry, the dust is mainly
composed of amorphous silicates, but also bands of crystalline silicates are
found. In the carbon dominated chemistry, the dust consists mainly of amorphous
carbon, SiC, and probably MgS
\citep[the carrier of the broad 30$\mu$m band,][]{hony02}. Also the
``Unidentified InfraRed'' (UIR) bands are seen in these IR spectra, commonly
attributed to Polycyclic Aromatic Hydrocarbons (PAHs). A sub-class of C-rich
post-AGB stars shows a feature at 21$\mu$m, discovered by
\citet{kwok89}, possibly attributed to TiC nanocrystals \citep{vonhelden00}.

Finally, there is an intriguing diversity in the observed {\em photospheric
chemical patterns} of the post-AGB stellar sample. The third dredge-up is
expected to bring carbon and freshly synthesized s-process elements to the
photosphere of an AGB star. The spectral lines of these elements are thus
expected to be strong in the spectra of post-AGB stars. This is indeed what we
observe, but remarkably only in a certain sub-class of post-AGB objects. There
are very similar objects (with a comparable mass and metallicity) that do not
show these enhancements. This chemical dichotomy is surprisingly strict in the
sense that an object is either severly enriched, or it is not enriched at all
(even s-process {\em deficiencies} are observed). This is illustrated in
Fig.~3 of \citet{vanwinckel03} on which the overabundance of the s-process
element zirconium (atomic number Z\,=\,40) is plotted against metallicity for
a sample of stars that all show similar characteristics for a genuine post-AGB
classification, like a double peaked SED, low metallicity and kinematics
pointing to membership of an old population. The chemical difference is
directly clear from the spectrum itself and we refer to Fig.~\ref{fig:rsmblnce}
for an illustration. Moreover, all s-enriched objects studied till now are
suspected or confirmed 21$\mu$m sources. Six of these sources were recently
analysed in a homogeneous abundance study by \citet{vanwinckel00}, hereafter
\citetalias{vanwinckel00}. In the non s-enriched objects \citep[see e.g.][ for
abundance analyses of these objects]{luck90, vanwinckel97}, no enrichments were found larger than the [Zr/Fe]\,$\sim$\,0.2 error level.

In the present study, we present the analysis of two newly discovered
s-process enriched post-AGB stars, \irnulzes\ and \irnulacht\
(Tab.~\ref{tab:datanlzsnlcht}). Despite the fact that these IRAS sources have
been both candidate post-AGB stars
for more than a decade \citep[as derived from their position in the IRAS
color-color diagram, e.g.][]{preitemartinez88}, they are both still poorly
studied. For \irnulzes\ this can partly be attributed to the optical weakness
of this source (Tab.~\ref{tab:datanlzsnlcht}), while for \irnulacht\ it is
less clear why this source has been neglected for such a long time. Based on
a low resolution spectrum (10.7\,\AA\,pixel$^{-1}$) taken with the 2.5\,m INT
on La Palma, \citet{slijkhuis92} attributed a spectral type F0I to the optical
counterpart of \irnulzes. In a very recent study, \citet{hrivnak03} found that
this is a too early spectral type and classified \irnulzes\ as a F5
supergiant. The spectral type of \irnulacht\ (F8I) was taken from
\citet{reddy96} and is based on a low resolution (5.7\,\AA\,pixel$^{-1}$)
spectrum taken with a 1\,m telescope.

The paper is organised as follows: In Sect. \ref{sect:observations} 
we describe the observations of the two programme stars, together with the
data reduction. In Sect. \ref{sect:radialvlcts} the radial velocities of the
two programme stars are compared with literature values, with the aim of
testing their possible binary nature. Sect. \ref{sect:abundancenlss} focusses
on the technical aspects of the analysis, including atmospheric parameter
determination, line selection and spectrum synthesis. Also our abundance
results are presented in this section. In the next section, we compare the
two program stars with the 21$\mu$m stars of \citetalias{vanwinckel00}.
It turns out that the two stars share the same chemical properties with the
21$\mu$m stars. Sect. \ref{sect:comparisontrnmdls} is devoted to a detailed
comparison with theoretical chemical AGB model predictions. In this section,
also the 21$\mu$m stars are compared with these chemical models. Finally, in
Sect. \ref{sect:conclusion} we summarize our most important findings.

\begin{table*}
\caption{Basic parameters of the two objects discussed in this study.}\label{tab:datanlzsnlcht}
\begin{center}
\begin{tabular}{cccccccccccc}
\hline
IRAS&\multicolumn{2}{c}{Equatorial}&\multicolumn{2}{c}{Galactic}&\multicolumn{2}{c}{Visual} & Spectral & \multicolumn{4}{c}{IRAS Fluxes (Jy)} \\
 &\multicolumn{2}{c}{coordinates}&\multicolumn{2}{c}{coordinates}&\multicolumn{2}{c}{magnitude} & Type & $f_{12}$ & $f_{25}$ & $f_{60}$ & $f_{100}$ \\
 & $\alpha_{2000}$ & $\delta_{2000}$ & l & b & m(b) & m(v) & & & & & \\
\hline
06530-0213 & 06 55 32.1 & $-$02 17 30 & 215.44 & $-$0.13 & 16.4$^a$ & 14.0$^a$ & F0I$^a$ & 6.11 & 27.41 & 15.05 & \phantom{$<$}4.10\\
      &            &             &        &         & 16.3$^b$ & 14.1$^b$ & F5I$^c$ & & & & \\
08143-4406 & 08 16 02.9 & $-$44 16 01 & 260.83 & $-$5.07 & 14.1$^b$ & 12.4$^b$ & F8I$^b$ & 0.60 & \phantom{2}9.26 &  \phantom{1}6.06 & $<$3.73\\
\hline
\multicolumn{12}{l}{\small Source SIMBAD, except: $^a$\,\citealt{slijkhuis92}, $^b$\,\citealt{reddy96}, $^c$\,\citealt{hrivnak03}}
\end{tabular}
\end{center}
\end{table*}

\section{Observations}\label{sect:observations}
High resolution, high signal-to-noise VLT+UVES spectra of the two programme
stars are taken in the framework of our ongoing program to study the
photospheric chemical composition of stars in their last stages of evolution
\citep[e.g. Paper I; ][]{reyniers01, reyniers03}. The two targets are part
of a larger sample of eleven post-AGB objects that were observed in service mode
during two periods \citep[04-07/2000 and 01-02/2001, see ][]{reyniers02a}.
The resolving power of these spectra varies between $\sim$55,000 and
$\sim$60,000. Some details about the observations are given in
Tab.~\ref{tab:observations}.

\begin{figure*}
\resizebox{\hsize}{!}{\rotatebox{-90}{{\includegraphics{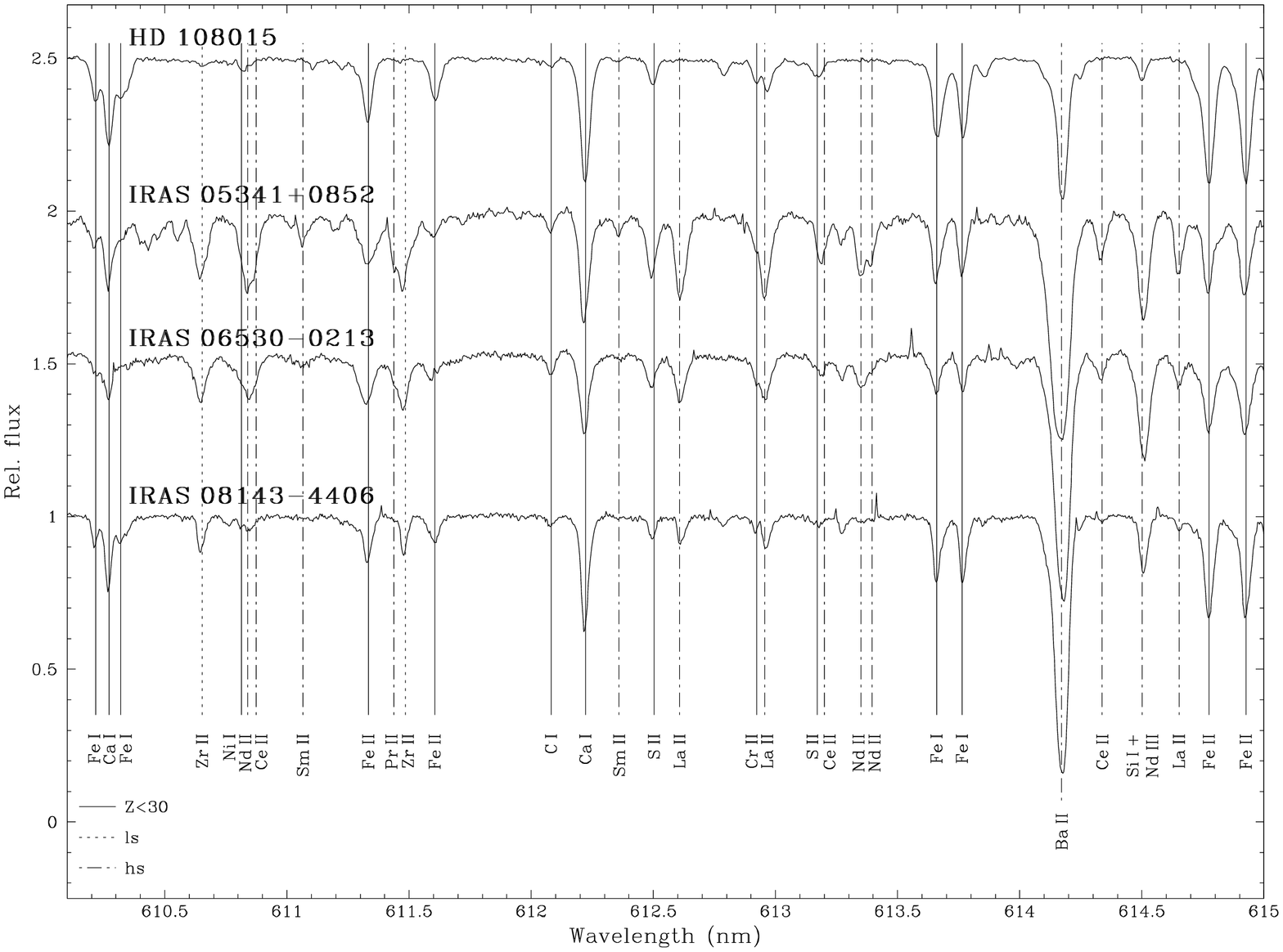}}}}
\caption{The spectra of \irnulzes\ and \irnulacht\ compared with the heavily
s-enriched post-AGB star \irnulvijf\ and the non s-enriched post-AGB star
\object{HD\,108015}. \irnulvijf\ is discussed in \citetalias{vanwinckel00},
and it is the most heavily s-process enriched star known so far.
\object{HD\,108015} has been analysed by \citet{vanwinckel97} and has
atmospheric parameters comparable with the other three ($T_{\rm eff}$,
$\log g$, $\xi_t$) = (7000\,K, 1.5, 4.0\,km\,s$^{-1}$) but is slightly less
metal deficient ([Fe/H]\,=\,$-$0.1). A complete line identification of this
spectral interval has been pursued by the use of the VALD database
\citep{kupka99}. Lines of light s-process elements (Sr peak) are identified
by a dotted line; lines of heavy s-process elements (Ba peak) by a dash-dotted
line; lines of other elements (which are mainly $\alpha$ and iron peak
elements) by a full line. The s-process enhancement of the three IRAS stars is
clear. \irnulvijf\ is the most s-enriched one, but the stronger lines are also
due to a slightly lower temperature of this object. All spectra are VLT+UVES
spectra, except the spectrum of \object{HD\,108015}, which is a ESO1.5+FEROS
spectrum taken at March 22, 2000.}\label{fig:rsmblnce}
\end{figure*}

The standard reduction was performed in the dedicated ``UVES context'' of the
MIDAS environment and included bias correction, cosmic hit correction,
flat-fielding, background correction and sky correction. We used optimal
extraction to convert frames from pixel-pixel to pixel-order space. The spectra
were normalised by dividing the individual orders by a smoothed spline function
defined through interactively identified continuum points. Orders were merged
after this normalisation. For a detailed description of the reduction
procedure, we refer to \citet{reyniers02a}. In Tab.~\ref{tab:observations}, we
also list some indicative signal-to-noise values of the final data product.
Sample spectra can be found in Figs.~\ref{fig:rsmblnce}, \ref{fig:xixmpl}
and \ref{fig:hyperfijnstrctr}.

\begin{table}\caption{Observational log. Spectral gaps occur between 577\,nm
and 583\,nm and between 854.4\,nm and 864.5\,nm due to the spatial gap between
the two UVES CCDs.}\label{tab:observations}
\begin{center}
\begin{tabular}{ccccc}
\hline
 date & UT    & exp.time & wavelength & S/N \\
      & start &  (sec)   & interval (nm) &     \\
\hline
\multicolumn{5}{c}{\irnulzes}\\
\hline
 2001-01-12 & 03:36 & 3$\times$1800 & 477.5$-$681 & 100 \\ 
\hline
\multicolumn{5}{c}{\irnulacht}\\
\hline
 2001-01-16 & 06:41 & 2$\times$1800 & 374.5$-$498 & 130 \\ 
 2001-01-16 & 06:41 & 1$\times$1800 & 670.5$-$1055  & {\em sat.}\\ 
 2001-01-16 & 07:16 & \phantom{1}3$\times$500 & 670.5$-$1055 & 200 \\ 
 2001-02-01 & 04:21 & 1$\times$1800 & 477.5$-$681 & 150 \\ 
\hline
\end{tabular}
\end{center}
{\em sat.} = saturated
\end{table}

\section{Radial velocities}\label{sect:radialvlcts}
Since binarity can influence the chemical evolution of a post-AGB star
drastically \citep[see e.g.][]{vanwinckel95}, it is important to determine and
monitor accurate radial velocities for these objects. The large wavelength
coverage together with the high resolution of our UVES spectra permit us to
determine these velocities using a large number of lines. The obtained radial
velocities are gathered in Tab.~\ref{tab:radialvlcts}, together with the few
values found in the literature. An additional velocity is obtained with the
CORALIE spectrograph mounted on the Swiss telescope Euler in La Silla, Chile.
Highly accurate velocities are obtained with this instrument by
cross-correlation observed spectra with a spectrum mask of comparable
spectral type.

It is obvious that there is still monitoring needed to draw any final
conclusion, but till now, there is no evidence for a binary motion for the two
programme stars.

\begin{table}
\caption{Heliocentric radial velocities of the programme stars, both from the
literature and from this study. If the velocity is obtained in this study, the
number of lines on which it is based is given in the last column; otherwise
the reference is given in the same column. The last velocity is obtained with
the CORALIE spectrograph. Typical errors on these velocities are between
1.0 and 1.5\,km\,s$^{-1}$.}\label{tab:radialvlcts}
\begin{center}
\begin{tabular}{cccc}
\hline
Date         & v$_r$        & method$^\dag$ & ref.$^\ddag$ or number\\
(yyyy-mm-dd) & (km\,s$^{-1}$) &          & of used lines\\
\hline
\multicolumn{4}{c}{\irnulzes}\\
\hline
1991-04\phantom{-11} & 50\phantom{.0} & a & ref. 1\\
1997-10-17   & 51.0  & b & ref. 2\\
2001-01-12   & 49.8  & b & n\,=\,236 \\
2001-12-10   & 50.4  & b & ref. 2\\
\hline
\multicolumn{4}{c}{\irnulacht}\\
\hline
2001-01-16   & 51.5  & b & n\,=\,134\\
2001-02-01   & 52.2  & b & n\,=\,434\\
2002-05-22   & 49.6  & c & \\
\hline
\end{tabular}
\end{center}
$^\dag${\em method} a: CO (J\,=\,2$-$1) line; b: mean of individual optical
lines c: cross-correlation with spectrum mask
$^\ddag${\em ref.} 1: \citealt{hu94}; 2: \citealt{hrivnak03}\\
\end{table}

\section{Abundance analysis}\label{sect:abundancenlss}
\subsection{Atomic data and atmospheric parameters}
A list of lines useful for the chemical analysis of A and F type stars has
been collected at the Instituut voor Sterrenkunde during the past few years
and is regularly updated. This list is described in detail in
\citetalias{vanwinckel00}. A recent update of the list after the publication
of \citetalias{vanwinckel00} consists of the inclusion of the data published
by \citet{lawler01a} for La and \citet{lawler01b} for Eu. Using this same list,
we do not only restrict ourselves to lines with reliable atomic data, but we
also ensure that the different analyses are perfectly homogeneous.

Equivalent widths of unblended lines were measured with direct integration;
multiple gaussian fitting was applied for lines with (a) blended wing(s). 
The entire line list, together with the measured equivalent widths for both
stars can be found in Tab.~\ref{tab:equivalentwdths}, which is published in
electronic form at the CDS. The atmospheric parameters were obtained by the
usual spectroscopic method in which the effective temperature is obtained by
imposing the iron abundance, derived from the individual Fe\,{\sc i} lines,
to be independent of lower excitation potential; the gravity by imposing
ionization equilibrium; the microturbulent velocity by imposing iron abundance
derived from the individual Fe\,{\sc i} lines to be independent of (reduced)
equivalent width. The model atmospheres of \citet{kurucz93} were used, in
combination with the latest version (April 2002) of the LTE abundance
calculation routine MOOG \citep{sneden73}.

\begin{table}
\caption{Line identifications and measured equivalent widths for \irnulzes\ 
and \irnulacht. This table is only available in electronic form at the
CDS. It contains the following columns: the rest wavelength (in \AA), the 
identification, the lower excitation potential (in eV), the adopted $\log gf$
value, the measured equivalent widths for \irnulzes\ (in m\AA), the measured
equivalent widths for the red setting (2001-02-01, see
Tab.~\ref{tab:observations}) of \irnulacht\ and the measured equivalent widths
for the dichroic setting (2001-01-16, see Tab.~\ref{tab:observations}) of
\irnulacht.}\label{tab:equivalentwdths}
\end{table}

\begin{figure}
\resizebox{\hsize}{!}{\includegraphics{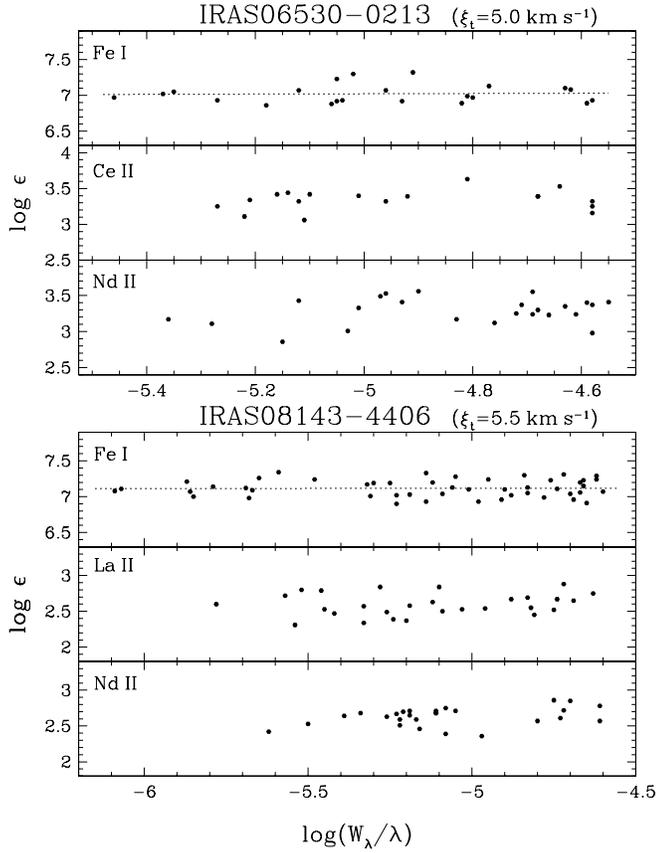}}
\caption{Reduced equivalent width - abundance diagrams for elements of which
the abundance is based on more than fifteen lines. Such diagrams are constructed
to derive the microturbulent velocity $\xi_{\rm t}$ for the model atmosphere;
the correct value for $\xi_{\rm t}$ is found if no trend is seen on the diagram.
In this figure, we check if the abundances of some non-iron species are
consistent with the value for $\xi_{\rm t}$ that was derived from a study of the
iron lines.}\label{fig:microturbrprt}
\end{figure}

The results on other species can be used as a check for the atmospheric
parameters, providing that the number of lines for that species is large
enough. Elements for which we detected lines of different ions can be used as
a first consistency check of the model atmosphere. A temperature check is
difficult since for most elements, the range in excitation potential is quite
small. Therefore, only the microturbulent velocity parameter $\xi_{\rm t}$ can
be checked in this way. For a given element, the most straightforward method is
then to search for that $\xi_{\rm t}$ for which the slope of the linear fit
on a reduced width - abundance diagram equals zero. With this method, however,
one loses the particular characteristics of the specific diagram, in the sense
that mostly outliers in $\log(W_{\lambda}/\lambda)$ determine the obtained
slope. Therefore, it is often much more instructive to inspect the diagram by
eye, than to rely on a least-squares method. This is done in
Fig.~\ref{fig:microturbrprt}, on which we plotted the Fe, Ce and Nd diagrams
for \irnulzes, and the Fe, La and Nd diagrams for \irnulacht. From this figure
it is clear that (a) the spread in reduced width $\log(W_{\lambda}/\lambda)$ is
significantly larger for Fe than for the other elements and that (b) the spread
in abundance $\log\epsilon$ is smaller in the case of iron. For the non-iron
species on Fig.~\ref{fig:microturbrprt} it is difficult to decide whether a
real trend is seen or not. Especially the points in the Nd diagram display a
somewhat confusing pattern. Several authors already stressed the need for new
atomic data for Nd\,{\sc ii} \citep[see e.g.][ for a discussion of the
available Nd\,{\sc ii} data]{sneden02}. We can conclude that there is no clear
indication for a dependence of $\xi_{\rm t}$ on the species that is used to
derive it. Note that an increase of $\sim$3\,km\,s$^{-1}$ does not induce large
changes in the derived abundances (a decrease of $\sim$0.1\,dex for the
strongest lines in our line list).

A particular problem arose in the determination of the parameters of
\irnulacht. Contrary to \irnulzes, the star is measured in
two different spectrograph settings: on 2001-01-16 the dichroic setting
(hereafter {\sf\small DI2}) is used with a simultaneous observation in
the blue and in the red, while on 2001-02-01 the red setting centered on
580\,nm (hereafter {\sf\small RED580}) is observed (see also
Tab.~\ref{tab:observations}).
Between these two observations, there is a time gap of 16 days. By inspecting
the spectral overlap of the two observations, we concluded that the atmospheric
parameters apparently changed during this time gap. The adoption of two
different sets of atmospheric parameters for each setting is therefore
necessary. Unfortunately, the number of Fe lines suitable for parameter
determination is too low for the dichroic setting. An alternative method
had to be developed to obtain the parameters for this setting.

We chose to deduce the atmospheric parameters for the 2001-01-16 setting
of \irnulacht\ {\em relative} to its parameters of the 2001-02-01 setting,
by demanding that the lines in the overlap should yield the same abundance.
So, after determination of the atmospheric parameters for \irnulacht\ for
the 2001-02-01 setting, we searched for the model parameters for the
2001-01-16 setting, by minimizing the abundance difference of the lines
in the overlap. This yielded model parameters ($T_{\rm eff}$, $\log g$,
$\xi_t$) = (7250\,K, 1.5, 5.5\,km\,s$^{-1}$) for the 2001-02-01 setting,
and (7050\,K, 1.2, 5.5\,km\,s$^{-1}$) for 2001-01-16. One should note
that a difference of 0.3\,dex in gravity is rather large: assuming a mass of
1.5\,M$_{\odot}$, a $\log g$ of 1.5 yields an approximative radius of
R\,$\simeq$\,36\,R$_{\odot}$, while a $\log g$ of 1.2 yields
R\,$\simeq$\,51\,R$_{\odot}$. This results in a pulsation velocity
of $\overline{\rm v}$\,$\simeq$\,7.5\,km\,s$^{-1}$, which is not consistent
with the seemingly constant radial velocity of \irnulacht\ (see
Tab.~\ref{tab:radialvlcts}). Moreover, such a change in atmospheric
parameters would induce a difference of $\sim$0.6\,mag in V in 16 days.
Since the gravity is not well constrained (a typical error in $\log g$ is
0.5), we decided to keep the difference in the parameters in order to keep
the abundances deduced from the two settings as consistent as possible.

\subsection{Spectrum synthesis of \irnulzes}
\subsubsection{The macroturbulent broadening $\xi_m$}
When computing a spectrum synthesis, a broadening parameter, additional to the
usual atmospheric parameters, is required to match the synthetic line profiles
with the observed ones. This broadening is the combined effect of instrumental,
macroturbulent and rotational broadening. The latter one is, however, thought
to be small in F-G supergiants. Since the instrumental broadening can be
deduced from the width of the Th-Ar lines of the calibration spectra, the
macroturbulent broadening ($\xi_m$) is the only parameter that has to be
estimated before comparing synthetic spectra with observed ones.
We concentrated on the spectra of \irnulzes, as this star is the most
interesting in the framework of s-process nucleosynthesis.

\begin{figure}
\resizebox{\hsize}{!}{\rotatebox{-90}{\includegraphics{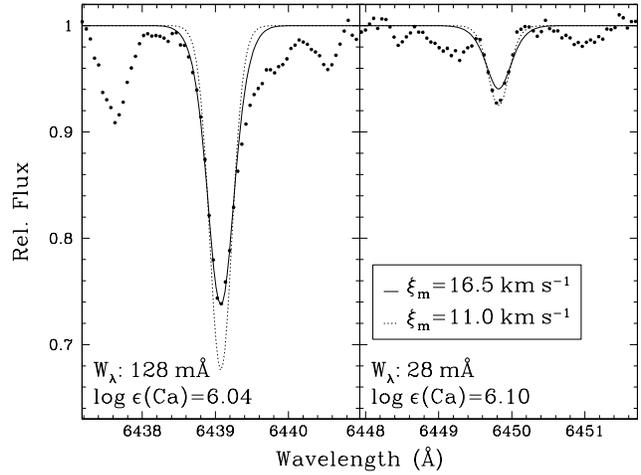}}}
\caption{Synthesis of two lines of neutral calcium. From this figure, it is
clear that spectral lines cannot be fitted using one fixed value for the
macroturbulent broadening $\xi_m$. The synthesis is made using the abundance
as derived from their equivalent width. The two abundances are in good
agreement. The best-fit macroturbulent broadening, however, differs by
5.5\,km\,s$^{-1}$. }\label{fig:xixmpl}
\end{figure}

For the instrumental broadening we took the median of the resolution of the
Th-Ar lines used in the calibration, which is
$\delta\lambda$\,=\,$\lambda$/59525\,\AA\ for the lower (EEV) CCD
and
$\delta\lambda$\,=\,$\lambda$/56575\,\AA\ for the upper (MIT) CCD.
The macroturbulent velocity $\xi_m$ was then obtained as follows.
We selected 121 unblended lines from the abundance analysis of \irnulzes\ 
from different species and ionisation (neutral as well as singly ionised). Each
of these 121 lines was synthesized with MOOG using the abundance as derived
by its equivalent width. The macroturbulent broadening $\xi_m$ is the only free
parameter in this synthesis. We fitted the synthetic spectra by varying
this parameter in steps of 0.5\,km\,s$^{-1}$. The profile used for the
macroturbulent broadening is a radial-tangential macroturbulence profile based
on the work of \citet{gray92}. During this fitting procedure, we soon realized
that we could not fit the lines with one fixed value for $\xi_m$, but that each
line required its own value. This is illustrated in Fig.~\ref{fig:xixmpl}, on
which we synthesized two calcium lines. These lines are lying in the same
spectral region and the abundances derived from their equivalent widths are in
very good agreement. However, we need two different values for $\xi_m$ in a
synthesis of their profiles.

\begin{figure}
\resizebox{\hsize}{!}{\includegraphics{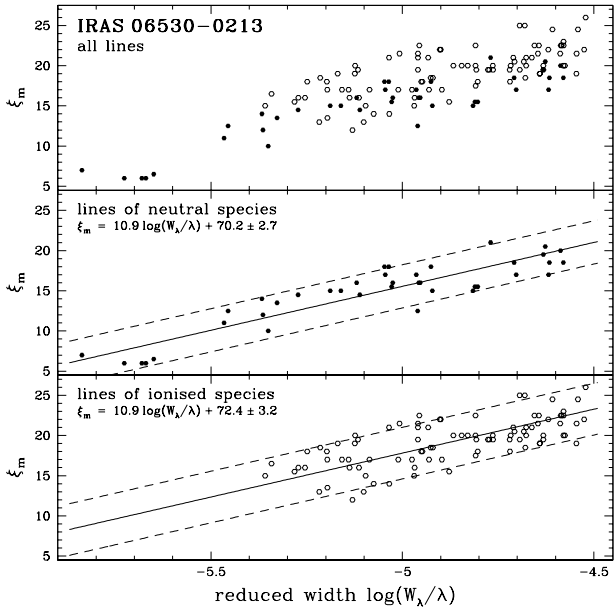}}
\caption{Relation between the (reduced) equivalent width
$\log({\rm W}_{\lambda}/{\lambda})$ and the macroturbulent velocity {$\xi_m$}
of 121 lines used in the analysis of \irnulzes. The lines from ionised species
have systematically higher macroturbulent velocities than lines from neutral
species of the same width {\em (upper panel)}. Therefore, two separate
relations are deduced ({\em middle} and {\em lower panel}). The errorbars
are the standard deviations on the mean trend multiplied by 1.5.}\label{fig:ximrltn}
\end{figure}

Analyzing these different values for $\xi_m$, we found a surprisingly tight
correlation of the macroturbulent velocity with the (reduced) equivalent width
(classical correlation coefficient $\rho$\,=\,0.81). A simple least squares fit
gives
\[ \xi_m = 10.9 \log({\rm W}_{\lambda}/{\lambda}) + 71.7 \]
Further inspection of the results gives an offset of $\sim$2.3\,km\,s$^{-1}$
between lines of neutral and lines of ionised species with the same reduced
width. Therefore, we can specify the previous relation by making the
distinction between lines of neutral and lines of ionised species:
\begin{eqnarray}
\xi_m({\rm neutral\ lines}) &=& 10.9 \log({\rm W}_{\lambda}/{\lambda}) + 70.2 \pm  2.7 \nonumber \\
\xi_m({\rm ionised\ lines}) &=& 10.9 \log({\rm W}_{\lambda}/{\lambda}) + 72.4 \pm  3.2 \nonumber 
\end{eqnarray}
where the errorbar is empirically deduced from Fig.~\ref{fig:ximrltn}
(being 1.5 times the standard deviation on the mean trend).
The correlation coefficient for lines of neutral species is
$\rho_{\rm n}$\,=\,0.90;
the same coefficient for lines of singly ionised species is
$\rho_{\rm i}$\,=\,0.70.
The reason for this dependence is not clear, nor a description of this effect
was found in literature. This might be due to the fact that most authors
deduce the macroturbulent velocity from just a few lines or lines in the
immediate proximity of the line/region under interest. We interpret this
effect as an optical depth effect of the macroturbulent velocity, the stronger
lines being formed on average at lower geometrical depth. We want to stress
that a similar relation, however less documented, was found for
\object{HD\,172481} \citep{reyniers01}, so it is certainly not an isolated
peculiarity of the spectrum of \irnulzes\ studied here.

\subsubsection{O synthesis}
Due to the high excitation potential of the oxygen lines, a reliable oxygen
abundance is always difficult to derive in this temperature-gravity domain,
and mostly only the oxygen triplet at 6155\AA\ is accessible. The situation is
even worse for heavily s-enriched objects like \irnulzes\ since the triplet is
significantly blended by lines of s-process species.  Important blends at the
position of the triplet are lines of Fe\,{\sc i} (6157.73\,\AA), Pr\,{\sc ii}
(6157.81\,\AA) and Nd\,{\sc ii} (6157.82\,\AA). Therefore, the only way to
obtain a reliable oxygen abundance in heavily enriched objects is by spectrum
synthesis. For the O lines themselves, we used the atomic data published by
\citet{biemont91}.

Furthermore, the relation between equivalent width and macroturbulent velocity
presented above causes a double difficulty when making the synthesis.
First, the relation is deduced for single, non-blended lines. Hence, it is 
not clear how the relation should be applied to blended lines.
Second, since the three lines are of different strength, each line requires
its own macroturbulent broadening, so that the lines of the triplet had to be
separately fitted. As a consequence, it is also difficult to show the
synthesis on one single figure. Therefore, the results are summarized in a
table (Tab.~\ref{tab:xgn06530}).

\begin{table}
\caption{Results of the spectrum synthesis of the O\,{\sc i} triplet of
\irnulzes.}\label{tab:xgn06530}
\begin{center}
\begin{tabular}{ccccc}
\hline
$\lambda$&$\chi$  &$\log gf$& adopted $\xi_m$& $\log \epsilon$ \\
 (\AA)   & (eV)   &        & (km\,s$^{-1})$ &                 \\
\hline
6155.971 & 10.74  & $-$0.674 & 13.5\phantom{5} ($\pm$3) & 8.58 ($\pm$0.06)\\
6156.778 & 10.74  & $-$0.453 & 15.25 ($\pm$3)           & 8.71 ($\pm$0.09)\\
6158.187 & 10.74  & $-$0.307 & 16.25 ($\pm$3)           & 8.62 ($\pm$0.05)\\
\cline{5-5}
         &        &        &                       & \rule[-0mm]{0mm}{4mm}{\em 8.64 \phantom{($\pm$0.05)}}\\
\hline
\end{tabular}
\end{center}
\end{table}

\subsection{Elements beyond the Ba peak}
Inspired by the detection of a hafnium (Hf, Z=72) line in the objects
\irnulzeven\ and \ireennegen\ as reported in \citetalias{vanwinckel00},
we initiated a systematic search for other elements beyond the Ba peak. Our
VLT+UVES spectra of the heavily enriched objects are ideally suited for this
search since they combine a large wavelength coverage with a high
signal-to-noise. Moreover, the release of a new database with atomic data of
the lanthanides (Z\,=\,57--71) at Mons University (B), the so called D.R.E.A.M.
project, is an invaluable supplement to the VALD database which was our only
source of atomic data of species in this mass range till now. The potential
of this new database was recently illustrated by \citet{reyniers02b}. Since it
also contains atomic data for doubly ionised lanthanides, we extended our
search to these ions. The details of our successful search are published in
a dedicated letter \citep{reyniers03}, while here we only report on the
results of the two programme stars. We have derived abundances of gadolinium
(Gd, Z=64), ytterbium (Yb, Z=70), lutetium (Lu, Z=71) and tungsten (W, Z=74)
for \irnulzes, and of Gd and Lu for \irnulacht.

\subsection{Hyperfine structure}
In \citetalias{vanwinckel00} we studied the influence of hyperfine splitting
(hfs) on the abundances of elements which are considered to be sensitive for
this effect. The general conclusion of this exercise was that the hfs effect
on the lines used in the abundance analysis is very small, due to the
relatively high effective temperature of the 21$\mu$m stars. This is only
true for weak lines, while for stronger lines an effect is indeed expected.
Such strong lines were, however, not used in the abundance determination.
Because several papers discussing hfs constants for different elements
recently appeared in literature, we decided to repeat this exercise for the
two program stars in this paper. The influence of hfs was determined by
calculating the equivalent width of a studied line, with and without hfs
decomposition. In this way, the exact profile of the line is eliminated, and
no broadening factors have to be applied. We focussed on the s-process
elements La, Eu and Lu. 

For La, hfs constants A and B were taken from \citet{lawler01a}. The La line
list of \irnulzes\ contains only one line with hfs constants for both levels
(5482.268\,\AA, W$_{\lambda}$\,=\,91.8\,m\AA). The effect of hfs on the
abundance is smaller than 0.01\,dex. The La line list of \irnulacht\ contains
eight lines for which there are hfs constants in \citet{lawler01a}. For seven
of these lines, the effect is negligible ($<$0.02\,dex), while for the
strongest line in the list (6262.287\,\AA, W$_{\lambda}$\,=\,148.0\,m\AA), the
effect is significant. An hfs-treatment of this line yields an abundance of
2.60\,dex, which is 0.15\,dex lower than a treatment without hfs.

The Eu abundance is often considered to be strongly hfs sensitive, and an hfs
treatment is indispensable when studying Eu abundances in cool stars. We
studied the strongest Eu line of the two stars that was used in our analysis
(6645.064\,\AA, 103.2\,m\AA\ in \irnulzes). A and B constants for this line
were taken from \citet{lawler01b}. The solar isotopic composition was applied.
It turned out that, even for this relatively strong line with a strong
decomposition, the effect on the derived abundance is marginal ($<$0.02\,dex).
We also studied the effect of hfs on the Eu abundance of the 21$\mu$m stars of
\citetalias{vanwinckel00}. Only for the cooler and strongly enriched
\irnulvijf, the effect was noticeable (an abundance decrease of 0.05\,dex and
0.07\,dex for the lines at 6437.640\,\AA\ and 7194.830\,\AA\ respectively).

\begin{figure}
\resizebox{\hsize}{!}{\rotatebox{-90}{\includegraphics{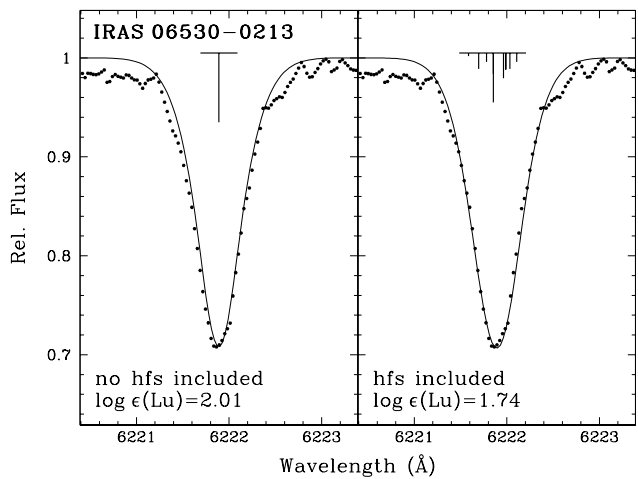}}}
\caption{Synthesis of the 6221.890\,\AA\ Lu line in \irnulzes, without
{\em (left panel)} and with {\em (right panel)} hyperfine decomposition. The
adopted abundance is indicated below. The difference in abundance illustrates
the necessity of the inclusion of hfs for this line.}\label{fig:hyperfijnstrctr}
\end{figure}

While for La and Eu, the effect of hfs on the abundance is much smaller than
other uncertainties ($\log gf$, continuum placement, undetected blends,
\ldots), the situation is different for the Lu abundance. Hfs constants for the
two detected Lu lines were taken from \citet{brix52} and \citet{denhartog98}.
In the case of \irnulzes, the difference in abundance between a hfs and a
non-hfs synthesis is $-$0.27\,dex for the 6221.890\,\AA\ line (see
Fig.~\ref{fig:hyperfijnstrctr}), and $-$0.14\,dex for the 6463.107\,\AA\ line.
For the less s-enhanced \irnulacht, hfs only affects the line profile of the
two Lu lines, but not their equivalent width.

\subsection{Abundance results}\label{subs:abundancerslts}
\begin{figure}
\resizebox{\hsize}{!}{\includegraphics{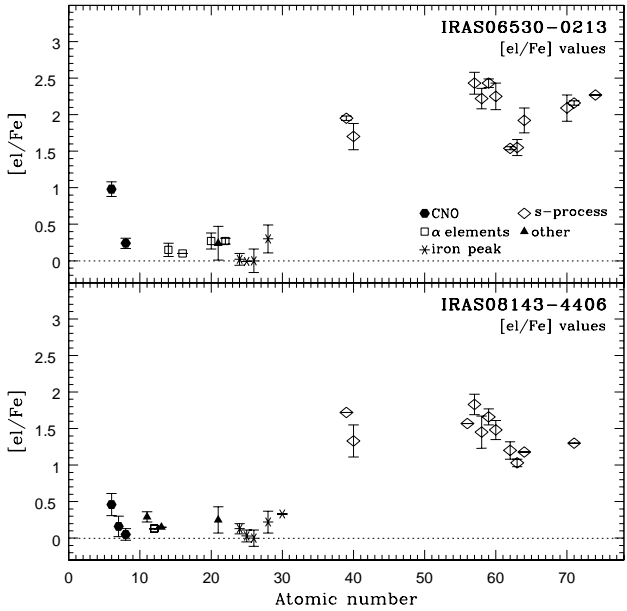}}
\caption{[el/Fe] values for \irnulzes\ and \irnulacht. The errorbars are the
line-to-line scatters; the dotted line represents [el/Fe]\,=\,0. It is
already clear from the figure that a higher neutron irradiation is expected
for \irnulzes.}\label{fig:elfewrdn0608}
\end{figure}

In Tab.~\ref{tab:a06a08sumout} we present the complete abundance analysis
results of \irnulzes\ and \irnulacht. The same results are graphically
presented in Fig.~\ref{fig:elfewrdn0608}. On this figure, the different groups
of elements are marked with different symbols. We will now summarize the
main results for each of these groups.
\begin{description}
\item{\em Metallicity}
Both stars are (moderately) metal deficient, with iron abundances of
[Fe/H]\,=\,$-$0.5 and $-$0.4 for \irnulzes\ and \irnulacht\ respectively.
The other iron peak elements all follow this deficiency, although zinc
(Zn) is somewhat higher than expected for \irnulacht.
\item{\em CNO-elements}
Both stars are clearly carbon enriched, with a huge enhancement in the case
of \irnulzes\ ([C/Fe]\,=\,$+$1.0). As a consequence, we derive a high C/O
number ratio of C/O\,=\,2.8 for this star; For \irnulacht, this ratio is
1.3. One has to note, however, that the errors on mainly the oxygen abundance
preclude an accurate C/O number ratio determination based on photospheric
atomic lines.
\item{\em $\alpha$-elements}
The simple mean of the [el/Fe] values of the (available) $\alpha$-elements
Mg, Si, S, Ca and Ti yields for both stars [$\alpha$/Fe] = $+$0.2. Such an
enhancement is normal for stars in this metallicity range, as a consequence
of the galactic chemical evolution and therefore does not correspond to an
intrinsic enhancement.
\item{\em s-process elements}
From Fig.~\ref{fig:elfewrdn0608}, it is clear that the s-process enrichment
of the two objects under study is very strong. This enrichment is the most
important argument for the post third dredge-up status of the two stars.
\end{description}
The s-process elements observed in evolved stars can be divided into two
groups: the light s-process elements around the magic neutron number 50 (Sr,
Y, Zr) and the heavy s-process elements around the magic neutron number 82
(Ba, La, Ce, Pr, Nd, Sm). In order to study the s-process pattern in more
detail, four indices are generally defined: [s/Fe], [ls/Fe], [hs/Fe] and
[hs/ls]. Which elements are taken into account to determine these indices
is different from author to author and is mainly determined by the
possibility to compute accurate abundances of the different elements. To be
able to compare our results with our results on the 21$\mu$m objects
\citepalias{vanwinckel00}, we define the ls-index as the mean of Y and Zr and
the hs-index as the mean of Ba, La, Nd and Sm. Consequently, [s/Fe] is the
mean of the former six elements and [hs/ls]=[hs/Fe]$-$[ls/Fe].

With the definitions from above, the s-process indices for \irnulzes\
are: [s/Fe]\,=\,$+$2.1, [ls/Fe]\,=\,$+$1.8, [hs/Fe]\,=\,$+$2.2 and
[hs/ls]\,=\,$+$0.4; for \irnulacht, they are: [s/Fe]\,=\,$+$1.5,
[ls/Fe]\,=\,$+$1.5, [hs/Fe]\,=\,$+$1.5 and [hs/ls]\,=\,0.0.
Note that the Ba abundance for \irnulzes\ had to be estimated in order to
calculate the [hs/Fe] and [hs/ls] indices. We estimated this abundance
in the same way as explained in \citetalias{vanwinckel00},
by using the tables of \citet{malaney87} and an exponential distribution of
neutron exposures of $\tau_0$\,=\,0.4\,mbar$^{-1}$. The Ba abundance estimated
from the La abundance is then [Ba/Fe]\,=\,$+$2.55.

These abundance analyses make \irnulzes\ and \irnulacht\ join the group
of post-AGB stars that clearly display chemical evidence for a third dredge-up
event. They share the same chemical signatures as the 21$\mu$m objects
discussed in \citetalias{vanwinckel00}. \irnulzes\ is s-process enhanced at
approximately the same level as \irnulvijf, the most s-process enriched
intrinsic object found so far.

\begin{table*}
\caption{Abundance results for the two programme stars.
Because of a time gap of 16 days between the {\sf\small DI2} and
{\sf\small RED580} UVES setting for \irnulacht, we had to adopt 2 different
sets of atmospheric parameters. The parameters for the {\sf\small RED580}
setting are deduced using the iron lines. The parameters for the
{\sf\small DI2} setting are deduced relative to the parameters of the
{\sf\small RED580} setting, by a detailed study of 26 lines in the
overlapping region. The following solar abundances are adopted: for the
solar iron abundance we used the meteoritic iron abundance of 7.51; the
references for the solar CNO are C: \citealt{biemont93}, N: \citealt{hibbert91}
and O: \citealt{biemont91}; for Mg and Si the latest Holweger values
\citep{holweger01} were used; for La and Eu we took the recent values derived
by \citet{lawler01a} and \citet{lawler01b} respectively; other solar abundances
were taken from the review by \citet{grevesse98}. Despite the fact that there
are more recent values for some of the solar abundances (especially for the
solar CNO), we take these references to ensure as much as possible the 
$gf$ values that we have used in the present paper to be consistent with the
adopted solar abundances.}\label{tab:a06a08sumout}
\begin{center}
\begin{tabular}{l|rrrrr|rrrrrrrrrrr|r} 
\hline
&
\multicolumn{5}{c|}{\rule[-0mm]{0mm}{5mm}{\large\bf \irnulzes}}&
\multicolumn{11}{c|}{\rule[-0mm]{0mm}{5mm}{\large\bf \irnulacht}}&\\
&
\multicolumn{5}{c|}{[Fe/H]\,$=$\,$-$0.46}&
\multicolumn{11}{c|}{[Fe/H]\,$=$\,$-$0.39}&\\
 & & & & & & \multicolumn{3}{c}{{\sf\small RED580} setting} & \multicolumn{3}{c}{{\sf\small DI2} setting} & \multicolumn{5}{c|}{final result} & \\
 & & & & & & \multicolumn{3}{c}{\em (2001/02/01)} & \multicolumn{3}{c}{\em (2001/01/16)} & & & & & \\
&
\multicolumn{5}{c|}{
$
\begin{array}{r@{\,=\,}l}
T_{\rm eff} & 7250\,{\rm K}  \\
\log g  & 1.0\ {\rm(cgs)} \\
\xi_{\rm t} & 5.0\ {\rm km\,s}^{-1} \\
\end{array}
$
}&
\multicolumn{3}{c}{
$
\begin{array}{r@{\,=\,}l}
T_{\rm eff} & 7250\,{\rm K}  \\
\log g  & 1.5\ {\rm(cgs)} \\
\xi_{\rm t} & 5.5\ {\rm km\,s}^{-1} \\
\end{array}
$
}&
\multicolumn{3}{c}{
$
\begin{array}{r@{\,=\,}l}
T_{\rm eff} & 7050\,{\rm K}\\
\log g  & 1.2\ {\rm(cgs)} \\
\xi_{\rm t} & 5.5\ {\rm km\,s}^{-1} \\
\end{array}
$
} & & & & & \\

\hline
ion & N &{\rule[0mm]{0mm}{4mm}$\overline{W_{\lambda}}$}&$\log\epsilon$&$\sigma$&[el/Fe]&
N&$\log\epsilon$&$\sigma$& N&$\log\epsilon$&$\sigma$& N & {\rule[0mm]{0mm}{4mm}$\overline{W_{\lambda}}$}
&$\log\epsilon$&$\sigma$& [el/Fe]& $\log\epsilon_{\odot}$\\
\hline
C\,{\sc i}    &  8 &  79 & 9.09 & 0.10 & 0.98   & 13 & 8.66 & 0.15 &  11 & 8.64 & 0.14 &  23 & 64 & 8.64 & 0.15 &    0.46 & 8.57 \\
N\,{\sc i}    &    &     &      &      &        &    &      &      &   7 & 7.76 & 0.14 &   7 & 62 & 7.76 & 0.14 &    0.16 & 7.99 \\
O\,{\sc i}    &  3 &{\em ss}&8.64&0.07 & 0.24   &  5 & 8.52 & 0.08 &     &      &      &   5 & 34 & 8.52 & 0.08 &    0.05 & 8.86 \\
\hline
Na\,{\sc i}   &    &     &      &      &        &  4 & 6.23 & 0.07 &     &      &      &   4 & 39 & 6.23 & 0.07 &    0.29 & 6.33 \\
Mg\,{\sc i}   &    &     &      &      &        &  1 & 7.28 &      &     &      &      &   1 & 40 & 7.28 &      &    0.13 & 7.54 \\
Al\,{\sc i}   &    &     &      &      &        &  1 & 6.38 &      &   2 & 6.22 &      &   2 & 10 & 6.23 &      &    0.15 & 6.47 \\
Si\,{\sc i}   &  1 &  43 & 7.29 &      & 0.21   & 11 & 7.58 & 0.16 &   3 & 7.46 & 0.15 &  13 & 28 & 7.57 & 0.15 &    0.42 & 7.54 \\
Si\,{\sc ii}  &  1 &  85 & 7.16 &      & 0.08   &    &      &      &     &      &      &     &    &      &      &         & 7.54 \\
S\,{\sc i}    &  1 &  44 & 6.97 &      & 0.10   &  3 & 7.12 & 0.01 &   6 & 7.07 & 0.11 &   6 & 55 & 7.07 & 0.11 &    0.13 & 7.33 \\
Ca\,{\sc i}   &  6 &  67 & 6.17 & 0.11 & 0.27   & 12 & 6.10 & 0.17 &     &      &      &  12 & 60 & 6.10 & 0.17 &    0.13 & 6.36 \\
Ca\,{\sc ii}  &    &     &      &      &        &  1 & 5.91 &      &     &      &      &   1 & 16 & 5.91 &      & $-$0.06 & 6.36 \\
Sc\,{\sc ii}  &  3 &  82 & 2.95 & 0.23 & 0.24   &  4 & 3.03 & 0.18 &     &      &      &   4 & 70 & 3.03 & 0.18 &    0.25 & 3.17 \\
Ti\,{\sc ii}  &  3 &  99 & 4.83 & 0.04 & 0.27   &  2 & 5.05 &      &     &      &      &   2 &127 & 5.05 &      &    0.42 & 5.02 \\
Cr\,{\sc i}   &  1 &  18 & 5.27 &      & 0.06   &  3 & 5.47 & 0.06 &   3 & 5.44 & 0.07 &   6 & 53 & 5.45 & 0.06 &    0.17 & 5.67 \\
Cr\,{\sc ii}  &  9 &  78 & 5.22 & 0.08 & 0.01   &  9 & 5.40 & 0.09 &   1 & 5.33 &      &   9 & 86 & 5.39 & 0.09 &    0.11 & 5.67 \\
Mn\,{\sc i}   &  1 &  14 & 4.92 &      &$-$0.01 &  2 & 5.03 &      &   2 & 5.06 &      &   3 & 59 & 5.03 & 0.08 &    0.03 & 5.39 \\
Fe\,{\sc i}   & 22 &  70 & 7.02 & 0.13 &$-$0.03 & 48 & 7.11 & 0.12 &   3 & 7.16 & 0.05 &  49 & 59 & 7.12 & 0.12 &         & 7.51 \\
Fe\,{\sc ii}  &  7 &  83 & 7.05 & 0.16 &        &  6 & 7.14 & 0.09 &   1 & 6.95 &      &   7 & 83 & 7.12 & 0.11 &         & 7.51 \\
Ni\,{\sc i}   &  4 &  22 & 6.09 & 0.19 & 0.30   & 13 & 6.10 & 0.17 &  10 & 6.07 & 0.12 &  19 & 41 & 6.08 & 0.15 &    0.22 & 6.25 \\
Zn\,{\sc i}   &    &     &      &      &        &  1 & 4.51 &      &   2 & 4.53 &      &   2 & 82 & 4.54 &      &    0.33 & 4.60 \\
\hline
Y\,{\sc ii}   &  2 &{\em bl}&3.73&0.03 & 1.95   &  1 & 3.58 &      &   1 & 3.55 &      &   2 &119 & 3.57 &      &    1.72 & 2.24 \\
Zr\,{\sc ii}  &  6 &  81 & 3.84 & 0.18 & 1.70   &  6 & 3.57 & 0.26 &   4 & 3.54 & 0.12 &   8 & 59 & 3.54 & 0.22 &    1.33 & 2.60 \\
\hline  	  
Ba\,{\sc ii}  &    &     &      &      &        &    &      &      &   1 & 3.31 &      &   1 & 33 & 3.31 &      &    1.57 & 2.13 \\
La\,{\sc ii}  & 13 &  77 & 3.10 & 0.15 & 2.43   & 28 & 2.59 & 0.16 &  13 & 2.55 & 0.09 &  37 & 62 & 2.57 & 0.14 &    1.83 & 1.13 \\
Ce\,{\sc ii}  & 18 &  74 & 3.34 & 0.14 & 2.22   & 14 & 2.59 & 0.21 &  10 & 2.70 & 0.24 &  24 & 55 & 2.64 & 0.22 &    1.45 & 1.58 \\
Pr\,{\sc ii}  &  8 &  84 & 2.68 & 0.06 & 2.43   &  9 & 1.98 & 0.11 &     &      &      &   9 & 40 & 1.98 & 0.11 &    1.66 & 0.71 \\
Pr\,{\sc iii} &  2 &  53 & 2.54 & 0.30 & 2.29   &    &      &      &     &      &      &     &    &      &      &         & 0.71 \\
Nd\,{\sc ii}  & 24 &  91 & 3.29 & 0.18 & 2.25   & 27 & 2.60 & 0.13 &   9 & 2.64 & 0.17 &  33 & 58 & 2.59 & 0.13 &    1.48 & 1.50 \\
Nd\,{\sc iii} &  2 & 124 & 3.05 & 0.01 & 2.01   &    &      &      &     &      &      &     &    &      &      &         & 1.50 \\
Sm\,{\sc ii}  &  2 &  34 & 2.09 & 0.02 & 1.54   &  3 & 1.88 & 0.16 &  17 & 1.81 & 0.12 &  18 & 61 & 1.82 & 0.12 &    1.20 & 1.01 \\
Eu\,{\sc ii}  &  2 &  72 & 1.61 & 0.11 & 1.55   &  2 & 1.18 &      &   1 & 1.14 &      &   3 & 32 & 1.16 & 0.05 &    1.03 & 0.52 \\
\hline
Gd\,{\sc ii}  &  3 &  41 & 2.58 & 0.17 & 1.92   &  2 & 1.91 &      &     &      &      &   2 & 10 & 1.91 &      &    1.18 & 1.12 \\
Yb\,{\sc ii}  &  2 &  47 & 2.71 & 0.18 & 2.09   &    &      &      &     &      &      &     &    &      &      &         & 1.08 \\
Lu\,{\sc ii}  &  2 &{\em ss}&1.76& 0.03 & 2.16  &  1 & 0.97 &      &     &      &      &   1 & 42 & 0.97 &      &    1.30 & 0.06 \\
W\,{\sc ii}   &  1 &{\em bl}&2.92&     & 2.27   &    &      &      &     &      &      &     &    &      &      &         & 1.11 \\
\hline
\multicolumn{17}{l}{{\em ss}: spectrum synthesis; {\em bl}: line is
blended, but abundance is corrected for this blend, using a dedicated MOOG
routine}\\
\end{tabular}
\end{center}
\end{table*}

\subsection{Comparison with \citet{hrivnak03}}
An independent abundance analysis of \irnulzes\ has recently been carried
out by \citet{hrivnak03}, based on high-resolution spectra taken with
the 2.7\,m telescope at McDonald Observatory. The authors derived a slightly
lower temperature of 6900\,K, resulting in a lower metallicity of
[Fe/H]\,=\,$-$0.9. The abundances relative to iron of the elements that are in
common in both analyses are, however, in good agreement. One should note that
the analysis presented here is based on spectra with both a higher resolution
(56\,000 vs. 45\,000) and a higher signal-to-noise (100 vs. 30). As a
consequence, our analysis does not only include more lines (149 vs. 102), it
also covers much more ions (28 vs. 17). 

\section{Neutron exposure}\label{sect:neutronxpsr}
In \citetalias{vanwinckel00} (Sect. 6) we compared the strength of the
neutron irradiation parameterized by the [hs/ls] index with both the total
s-process enrichment and the metallicity for the 21$\mu$m stars. A surprising
result from this comparison was the strong correlation between the [hs/ls]
index and the [s/Fe] index implying an increase in efficiency of the neutron
nucleosynthesis with increasing third dredge-up efficiency. This relation was,
however, hampered by the low number statistics on which it was based: only
six stars. Here we have two more data points (Tab.~\ref{tab:sfehslsfehslsan}),
one of which having a very strong enrichment (\irnulzes).

\begin{figure}
\resizebox{\hsize}{!}{\includegraphics{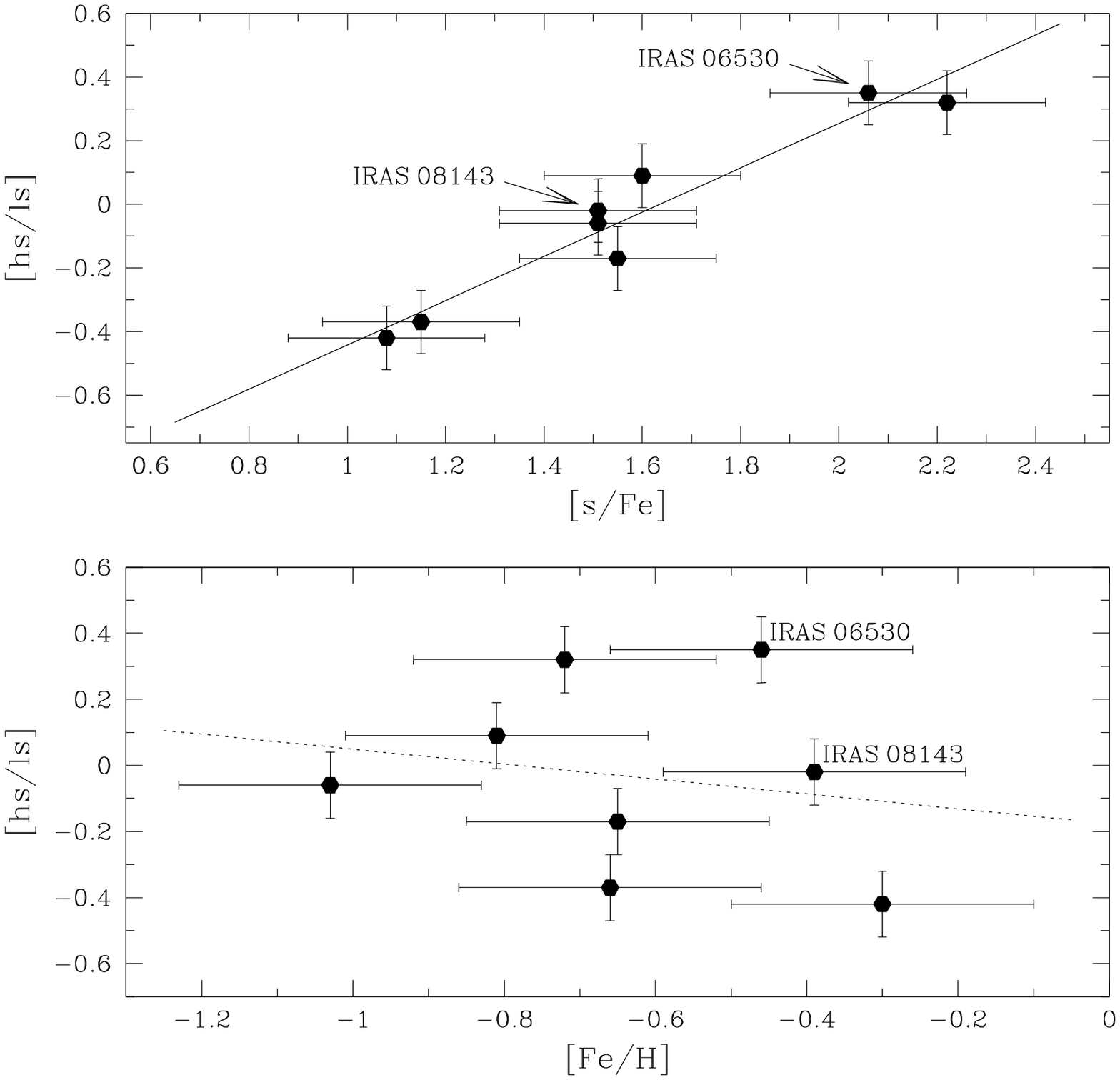}}
\caption{Comparison of \irnulzes\ and \irnulacht\ with the 21$\mu$m stars
of \citetalias{vanwinckel00}. The values of the data points in this figure
are given in Tab.~\ref{tab:sfehslsfehslsan}.
{\em upper panel:} The correlation between the total enrichment of
s-process elements ([s/Fe], which is the mean of [Y/Fe], [Zr/Fe], [Ba/Fe],
[La/Fe], [Nd/Fe] and [Sm/Fe]) and the [hs/ls] index. The straight line gives
the simple least-squares fit. {\em lower panel:} The [hs/ls] index
as a function of the metallicity determined by the Fe
abundance.  The dotted line gives the simple least-squares fit
[hs/ls]\,$=$\,$-$0.23\,[Fe/H]\,$-$\,0.18.}\label{fig:sfehslsfehslsan} 
\end{figure}

From Fig.~\ref{fig:sfehslsfehslsan} it is clear that the results of the two
newly discovered objects strengthen the conclusions of
\citetalias{vanwinckel00}.
In the {\em upper panel} of this figure,
the two data points for the newly discovered enriched stars nicely fit the
correlation between the [hs/ls] index and the [s/Fe] index. The new correlation
coefficient for the eight stars is +0.96, the same as found in
\citetalias{vanwinckel00}, while a least-squares fit gives
$[\mathrm{hs}/\mathrm{ls}] = 0.70 [\mathrm{s}/\mathrm{Fe}] - 1.14$.
Especially the analysis of \irnulzes\ yields an important data point to extend
the relation towards stronger enrichments. In turn, the addition of
the new objects dramatically illustrates the weak correlation between the
metallicity and the [hs/ls] index ({\em lower panel} of
Fig.~\ref{fig:sfehslsfehslsan}). Whereas the correlation coefficient for these
two quantities for the 21$\mu$m stars was still $-$0.55, it decreases to
$-$0.19 after addition of the two new objects. Mainly the strong enrichment of
\irnulzes\ compared to its only mild metal deficiency is a strong indication
to suspect a large {\em intrinsic} spread of integrated neutron irradiations
(or, alternatively, $^{13}$C pocket efficiencies). One can even doubt the
presence of any relation between [Fe/H] and [hs/ls] after inspection of
Fig.~\ref{fig:sfehslsfehslsan}.

\begin{table}
\caption{The metallicity and the s-process indices of the eight stars plotted
in Fig.~\ref{fig:sfehslsfehslsan}. Some values given in this table slightly
differ from the values given in \citetalias{vanwinckel00}, due to minor
revisions and updates that are described in detail in \citet{reyniers02a}.}\label{tab:sfehslsfehslsan}
\begin{tabular}{lccccc}
\hline
\multicolumn{1}{c}{object} & [Fe/H] & [s/Fe] & [ls/Fe] & [hs/Fe] & [hs/ls] \\
\hline
\irnulvier & $-$0.6 & 1.5 & 1.7 & 1.5 & $-$0.2 \\
\irnulvijf & $-$0.7 & 2.2 & 2.0 & 2.3 & \phantom{$-$}0.3\\
\irnulzes & $-$0.5 & 2.1 & 1.8 & 2.2 & \phantom{$-$}0.4\\
\irnulzeven & $-$1.0 & 1.5 & 1.6 & 1.5 & $-$0.1\\
\irnulacht & $-$0.4 & 1.5 & 1.5 & 1.5 & \phantom{$-$}0.0\\
\ireennegen & $-$0.7 & 1.2 & 1.4 & 1.0 & $-$0.4\\
\irtweetwee & $-$0.3 & 1.1 & 1.4 & 0.9 & $-$0.4\\
\irtweedrie & $-$0.8 & 1.6 & 1.5 & 1.6 & \phantom{$-$}0.1\\
\hline
\end{tabular}
\end{table}

\begin{figure}
\resizebox{\hsize}{!}{\includegraphics{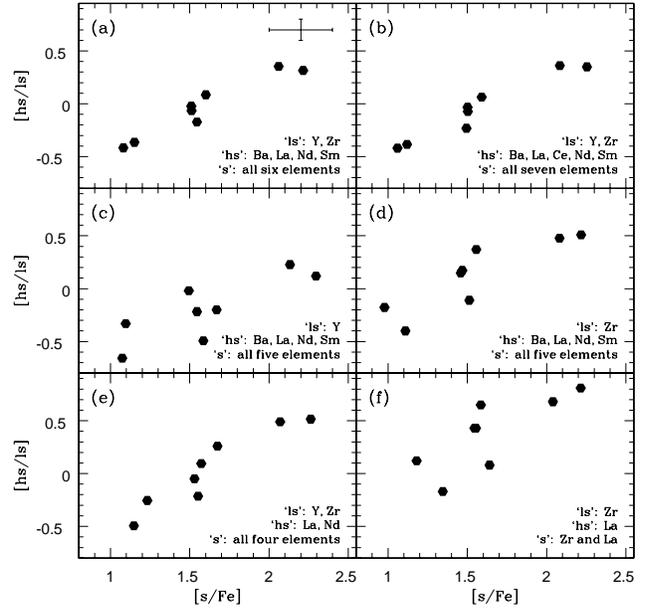}}
\caption{Study of the dependence of the proposed relation between the
[s/Fe] index and [hs/ls] index upon the particular definition of the
indices. Despite the scatter on the mean trend for some 
definitions, the relation holds for each definition presented here.
Especially the definition proposed by \citet{busso01} (panel e)
seems to be an excellent alternative for the ``default'' definition of
panel a.}\label{fig:smcorrdtcrtssm}
\end{figure}

The correlation between [hs/ls] and [s/Fe] is obviously dependent on the
specific choice of the elements involved in the calculation of both the
[s/Fe] and [hs/ls] index. The choice to incorporate Y and Zr for the
calculation of [ls/Fe] and Ba, La, Nd and Sm for [hs/Fe] was based on the choice made by \citet{busso95}. Some of these elements (like Ba and Sm) are,
however, not always observed and abundances of unobserved elements were
estimated using the \citet{malaney87} tables (see Sect.
\ref{subs:abundancerslts}). It is an interesting exercise to investigate a
possible dependence of the relation presented above upon the specific elements
used in the indices. On Fig.~\ref{fig:smcorrdtcrtssm}, the relation between
the neutron irradiation and the total s-process enrichment is revisited with
different elements for the indices.

On panel (a) of this figure, the relation is drawn for the ``default'' index
definition. On panel (b), the Ce abundance is incorporated in the `hs' (and
hence `s') definition, but the points hardly change their position on the
graph. On the {\em middle panels} the `ls' index is defined by only one
element, being Y on panel (c) and Zr on panel (d). This change has a rather
large impact on the position of the points. The restriction to one
abundance for the definition of the `ls' index does not seem favourable in this
context. Especially if the `ls' index is solely defined by the Y abundance,
the relation is clearly less well defined. In Sect.
\ref{sect:comparisontrnmdls} we further focus on the problems with this
particular abundance. On the two {\em lower panels}, the `hs' definition has
been varied. On panel (e), the [hs/ls] definition is the same as in
\citet{busso01}. With this definition, only {\em observed} abundances are
used, none were estimated. Moreover, the relation found in panel (a) is very
well reproduced with this definition. Hence, restricting hs to La and Nd, the
definition seems to be a very good alternative for the ``default'' definition.
In the last panel, panel (f), a ``minimal'' definition is given with only two
elements, Zr and La, which makes the scatter on the mean trend to become
quite high, but still a strong hint to the proposed relation exists.

\section{Comparison with AGB nucleosynthetic models}\label{sect:comparisontrnmdls}
\subsection{AGB stellar models}
Although there is general consensus that the s-process nucleosynthesis is
triggered by the engulfment of a small amount of protons of the hydrogen-rich
envelope into the top layers of the He intershell
\citep[e.g.][ and references therein]{busso99}, there is not yet a
selfconsistent nucleosynthetic AGB model. Different evolutionary codes use
different assumptions, not only for the partial mixing zone, but also for
the third dredge-up efficiency with or without overshoot assumptions
\citep[see comparative study of][ and references therein]{lugaro03}. 
Rotational mixing has recently been analysed and a strong effect seems
to be a decrease in the neutron irradiation since the major neutron
poison ($^{14}$N) is mixed into the $^{13}$C-rich layers
\citep{langer99, herwig03}.

In this section we compare our observed s-process distributions with the
results of nucleosynthetic AGB models
\citep{gallino98, busso99, busso01}. The nucleosynthesis predictions
are calculated by a postprocessing code grafted upon the FRANEC
evolutionary models \citep{straniero97, straniero03} that span a grid of
models with masses of 1.3 to 3.0 M$_{\odot}$ and metallicities from solar to
1/20 solar. In the models, an ad-hoc $^{13}$C profile is introduced which 
follows basically an exponential distribution \citep[see Fig.1 of][]{gallino98}
and which is burnt in radiative conditions during the interpulse phase,
at a temperature of $T$\,$\sim$\,0.9$\times$10$^8$\,K.
Although the physical origin of the $^{13}$C pocket remains to be studied,
the models are used here to estimate the strength of the $^{13}$C pocket
needed to explain the observed s-process abundances. The $^{13}$C pocket
choice labeled ``Standard Case'' (ST) corresponds to
4$\times$10$^{-6}$\,M$_{\odot}$ of $^{13}$C and it is named in this way because
it has been shown to best reproduce the ``main component'' of the s-process
in the solar system when applied to low mass AGB stars of [Fe/H]\,=\,$-$0.3
\citep{arlandini99}. Other choices of $^{13}$C pockets for individual AGB
stars are then obtained by multiplication or division of this standard
$^{13}$C amount. As a consequence, different s-process patterns can be
obtained for a fixed metallicity and mass. A spread in [hs/ls] such as found
in the lower panel of Fig.~\ref{fig:sfehslsfehslsan} is therefore naturally
found in these models when varying the strength of the $^{13}$C pocket.

The FRANEC models allow the third dredge-up mechanism to self-consistently 
occur after a limited number of pulses, using the Schwarzschild criterion
and without invoking any extra-mixing. Other AGB models
\citep[e.g.][]{goriely00, herwig03} use different prescriptions. Note that in
the FRANEC models the thermal pulses with third dredge-up stop prior to the
end of the AGB evolution, when the envelope is still relatively massive and
the superwind has yet to develop. The mass loss prescription is also a
critical point in each AGB model code, and different groups use different
approaches \citep[see e.g. ][ and references therein]{blocker99}. The models
presented here have been obtained using the mass loss parameterisation by
\citet{reimers75}, with the choice $\eta$\,=\,0.3 for the free parameter.

It is clear that there is not yet a full theoretical understanding of the AGB
nucleosynthesis and third dredge-up phenomena, making the matching between
the models and the observed s-process distributions to be interpreted with
caution. We want to stress, however, that the s-process enriched post-AGB
objects are ideally suited to confront observed s-process distributions with
theoretically predicted abundances, since (i) no dilution has to be accounted
for (as in the case of extrinsic AGB-stars) and (ii) these objects fully
completed their evolution through the TP-AGB phase (contrary to genuine AGB
stars from the M-MS-S-SC-C sequence).

\subsection{Model fits for the 21$\mu$m stars of \citetalias{vanwinckel00}}
We made an attempt to fit the s-process distribution of the 21$\mu$m stars
discussed in \citetalias{vanwinckel00} with the latest AGB models.
The abundances were re-calculated using the latest MOOG version (April 2002).
The abundances of \irnulvijf\ were re-determined using new VLT+UVES spectra.
Although this new analysis is based on superior spectra (see
Fig.~\ref{fig:rsmblnce}) compared to the WHT+UES spectra in
\citetalias{vanwinckel00}, the new results are not drastically different from
our previous analysis.

\begin{figure}
\resizebox{\hsize}{!}{\includegraphics{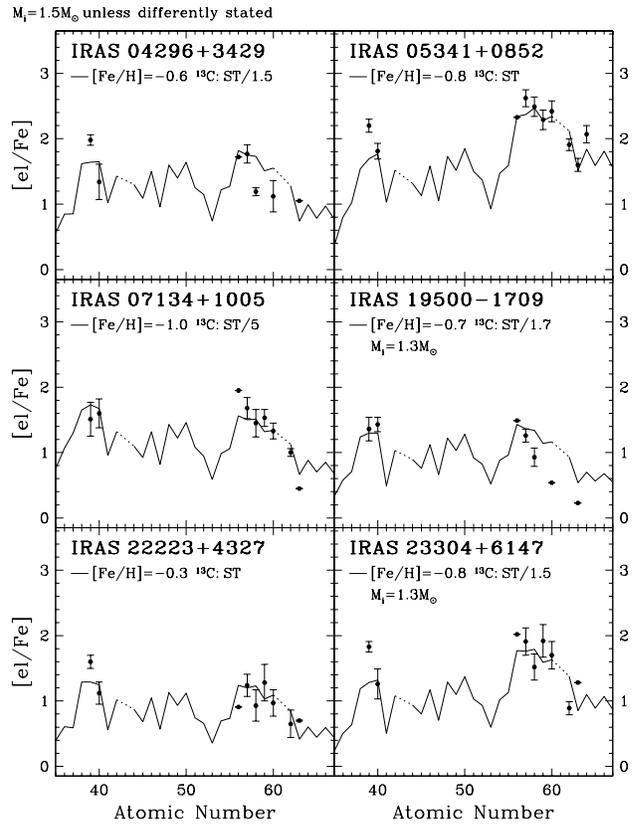}}
\caption{Abundances of the s-process elements for the six 21$\mu$m stars of
\citetalias{vanwinckel00}, together with their best fit AGB model. The
model parameters are given in the upper left corner; the initial mass is
taken 1.5\,M$_{\odot}$ for all models, except for \ireennegen\ and \irtweedrie.
For these two stars, a lower initial mass of 1.3\,M$_{\odot}$ yielded a better
fit. The errorbars plotted on the figure are the line-to-line
scatters. The actual error on an abundance is, however, in most cases
considerably larger than this value, due to uncertainties in the atmospheric
parameters, undetected blends, etc. Note that in the AGB model predictions
the solar meteoritic abundances by \citet{anders89} are used to convert the
s-process yields to the actual [el/Fe] predictions. This specific choice has
only a marginal effect on the predictions in this elemental range. Choosing the
solar abundances as listed in Tab.~\ref{tab:a06a08sumout} to calculate the
predictions, would introduce a maximum change of only 0.07\,dex (Pr) in the
predictions.}\label{fig:gallinofts}
\end{figure}

The model parameters were found in the following way.
We restricted the models to an initial mass of 1.5\,M$_{\odot}$. For the
metallicity of the model, we used the spectroscopically derived value. 
The $^{13}$C pocket is then the only parameter that is left over. It was found
by manually fitting the observed Zr and La abundance (which is in fact
representative for the [hs/ls] index). For two stars in the sample
(\ireennegen\ and \irtweedrie) no solution could be found. For these two stars,
the best fit was obtained with a model of lower initial mass of
1.3\,M$_{\odot}$. The fits are presented in Fig.~\ref{fig:gallinofts}.
The fit for \ireennegen\ still shows quite large discrepancies for the heavier
elements. The abundances of these elements are, however, derived from only a
few lines \citepalias[see][ for details]{vanwinckel00}, and should therefore
be considered with caution.

The fits are not unique, but it turned out that a free choice between the
possible $^{13}$C pockets was rather well constrained due to the fixed
metallicity. On the other hand, the amount of thermal pulses {\em can}
influence the choice of the $^{13}$C pocket: e.g. \citet{busso01} propose a
$^{13}$C pocket ``ST/3'' with 10 thermal pulses for the same abundances of
\irnulzeven\ we present here, whereas we found ``ST/5''. However, note that
the present AGB model predictions have been considerably improved since the
\citeauthor{busso01} paper. 

Finally, a small remark is given on the niobium (Nb, Z\,=\,41) abundance.
Since these models are for {\em intrinsic} stars, $^{93}$Zr with a half-life
of 1.5$\times$10$^6$\,yr, has not yet decayed to Nb. In {\em extrinsic}
stars, the Nb abundance is expected to be at the same level as Zr.

\subsection{Model fits for the two programme stars}
\begin{figure}
\resizebox{\hsize}{!}{\includegraphics{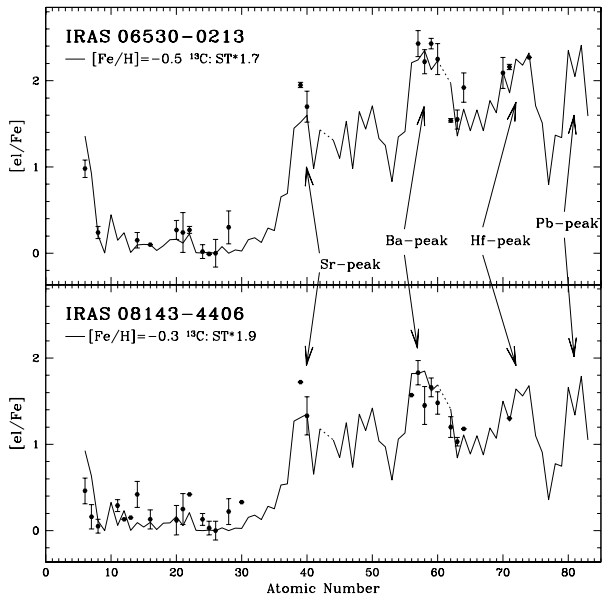}}
\caption{Abundances of \irnulzes\ and \irnulacht\ 
together with their best fit AGB model. The model parameters are given in
the upper left corner. The initial mass is 1.5\,M$_{\odot}$. Errorbars are
the line-to-line scatters that are given in Tab.~\ref{tab:a06a08sumout}.
Note that the use of the solar meteoritic abundances in the AGB model
predictions (see also the caption of Fig.~\ref{fig:gallinofts}), introduces
a rather large uncertainty for the W abundance, for which the difference
between solar photospheric and meteoritic abundance is very large
\citep[0.43\,dex,][]{anders89}.}\label{fig:gallinoftsfll}
\end{figure}

After this first exercise, we searched for the best model fit for the two
programme stars of this paper. The same fitting method was applied as for the
21$\mu$m stars, except that we extended the comparison to all observed
elements. The final model fits, together with their parameters, are shown in
Fig.~\ref{fig:gallinoftsfll}. Note that decreasing the metallicity by
e.g. 0.3\,dex
\citep[which corresponds to the metallicity found for \irnulzes\ by][]{hrivnak03}
would result in a fit with a $^{13}$C pocket which is decreased
with the same factor, with respect to the original one.

\subsection{Model fit conclusions}
Apart from the fact that several solutions are possible for the same pattern,
it is clear, though, that only one $^{13}$C pocket alone cannot explain the
abundance patterns in all eight objects. \irnulzeven\ for example can certainly
not be fitted with a $^{13}$C pocket which exceeds ``ST/2'', whereas for
\irnulvijf\ a small pocket ($<$``ST'') is not possible. This difference in
$^{13}$C pockets is much larger than the difference that could result from
uncertainties in the models or in the abundances, and therefore we believe
that the spread in $^{13}$C pockets is real. The need of an intrinsic
spread in the adopted $^{13}$C pockets should be regarded as an essential
result of this first exercise.

Other interesting conclusions can be drawn from the fits by inspecting the
abundances individually. We give four examples to illustrate the mutual
benefit of such fits for both the models and the observational analysis.\\
(1) The {\em carbon abundance} (not on the figure for the 21$\mu$m stars)
is predicted far too high in each fit. Since this particular abundance is very
reliable for each star ($\sigma$\,$<$\,0.2), the too high prediction is
probably a model's artefact.\\
(2) The {\em yttrium abundance} (Z=39) is systematically 0.2\,dex too high for
the cooler objects ($T_{\rm eff}$\,$\la$\,7000\,K) in the sample when compared
to the predicted abundances. For these cooler stars, only a few lines
(typically $\sim$4) were found since most Y\,{\sc ii} lines are too strong to
deduce an accurate abundance. The two hotter stars in the sample
(\irnulzeven\ and \ireennegen) do not show this effect, and since the hfs
effect for Y is only very small, this discrepancy between the observed and the
predicted abundance for the cooler stars is probably due to a $\log(gf)$
problem and/or blending of the Y\,{\sc ii}-lines used in these stars.\\
(3) Another, although minor, discrepancy is that the {\em cerium abundance} is
systematically too low compared to the predictions for at least five objects.
In this context it is noteworthy that the Ce abundance decreases by another
$\sim$0.2\,dex when calculated by using the $\log(gf)$ values of the D.R.E.A.M.
database. Hence, replacing the VALD data by the D.R.E.A.M. data for Ce would
only {\em increase} the observed discrepancy.\\
(4) The predictions for the elements beyond the Ba-peak for \irnulzes\ and
\irnulacht\  are very consistent with the observed abundances. This is an
additional argument strengthening the line identification for these elements.
Moreover, their abundances are very sensitive to the adopted $^{13}$C pocket.
Hence, these elements
(especially Hf and W) are ideally suited to discriminate between possible
$^{13}$C pockets. Unfortunately, a Hf abundance is very difficult to derive in
these objects since it has only suitable lines in the blue, and the W abundance
derived from the 5104.432\,\AA\ line depends on the adopted Sm abundance
\citep[see ][]{reyniers03}.

\section{Conclusion}\label{sect:conclusion}
\irnulzes\ and \irnulacht\ were classified as post-AGB candidates on the
base of their IR properties and their spectral type. The analysis presented
in this paper proved that both objects also show clear {\em chemical}
evidence of their evolved, post third dredge-up character, since they show not
only a clear carbon enhancement, but also a large enrichment in s-process
elements. Especially \irnulzes\ is an object that is very interesting for
further research since it is enhanced at almost the same level as \irnulvijf,
the most s-process enriched object known so far. Moreover, detailed abundances
of elements beyond the Ba peak were obtained, a result which was possible due
to the combination of the high quality VLT+UVES spectra and newly released
line data in both the VALD and the D.R.E.A.M. databases. The end point of the
s-process (Pb) is expected to be produced at important levels, also in these
moderate metal-deficient objects, but Pb has no suitable lines in the optical
spectrum for these temperatures and gravities.

The atmospheric and chemical properties of the two objects are remarkably
similar than those of the 21$\mu$m objects. Especially the relation between
third dredge-up efficiency and neutron nucleosynthesis efficiency outlined
in \citetalias{vanwinckel00} for the 21$\mu$m stars is confirmed
by these two new objects. On the other hand, the results on the two
objects made the relation between metallicity and neutron nucleosynthesis
efficiency even more confusing: whereas the 21$\mu$m stars of
\citetalias{vanwinckel00} still displayed a weak correlation, the new
data points (especially \irnulzes) suggest that there is no correlation
at all. This means that there is a large intrinsic spread in neutron
nucleosynthesis efficiency. Such a spread points to different $^{13}$C pocket
strengths in probably highly similar objects which is confirmed by our
modelling, in which quite different $^{13}$C pockets are needed to model the
s-process distribution. A physical explanation for the different $^{13}$C
pocket strengths is not yet found. The proton engulfment induces an interplay
between the $^{14}$N pocket, acting as a neutron poison, and the $^{13}$C
pocket, acting as a neutron donor. In this paper we restrengthen the
observational finding that even for objects with similar metallicities, quite
a different neutron irradiation is observed. Possibly rotation can play an
important role in this discussion \citep{langer99, herwig03}.

Moreover, apart from the s-process enriched objects discussed in this paper,
there are very similar post-AGB objects without showing any s-process
enrichment. This dichotomy between s-process enriched post-AGB stars, and
post-AGB stars showing no s-enrichment at all, is also not understood. Note,
however, that in the FRANEC models the third dredge-up does not occur below
a certain critical mass, which slightly decreases with decreasing metallicity.

It is clear that a spectroscopic infrared study of the two objects discussed
in this paper would be an invaluable supplement to this optical study. Indeed,
till now all enriched objects display the 21$\mu$m feature in their IR
spectrum if this data is available. It should be tested if this assertion
still holds after observing \irnulzes\ and \irnulacht\ in the IR spectral
domain.

\begin{acknowledgements}
MR and HVW acknowledge financial support from the Fund for Scientific Research
- Flanders (Belgium). This research has made use of the Vienna Atomic Line
Database (VALD), operated at Vienna, Austria, and the Database on Rare Earths
At Mons University, operated at Mons, Belgium.
\end{acknowledgements}


\end{document}